\documentclass[%
 reprint,
superscriptaddress,
 amsmath,amssymb,
 aps, prl,
floatfix,
]{revtex4-1}
\usepackage[normalem]{ulem}
\UseRawInputEncoding
\usepackage{graphicx}% Include figure files
\usepackage{color}
\usepackage{booktabs}
\usepackage{makecell}
\usepackage{dcolumn}% Align table columns on decimal point
\usepackage{bm}% bold math
\begin{document}
\bibliographystyle{apsrev4-1}
\preprint{APS/123-QED}

\title{Polarity control by inversion domain suppression in N-polar III-nitride heterostructures}

\author{Hengfang Zhang}
\affiliation{Center for III-Nitride Technology, C3NiT-Janz\'en and Department of Physics, Chemistry and Biology (IFM), Link\"{o}ping University, 581 83 Link\"{o}ping, Sweden}

\author{Ingemar Persson}
\affiliation{Thin Film Physics, Department of Physics, Chemistry and Biology (IFM), Link\"{o}ping University, 581 83 Link\"{o}ping, Sweden}

\author{Jr.-Tai Chen}
\affiliation{Center for III-Nitride Technology, C3NiT-Janz\'en and Department of Physics, Chemistry and Biology (IFM), Link\"{o}ping University, 581 83 Link\"{o}ping, Sweden}
\affiliation{SweGaN AB, Olaus Magnus v\"ag 48A, 583 30 Link\"oping, Sweden}

\author{Alexis Papamichail}
\affiliation{Center for III-Nitride Technology, C3NiT-Janz\'en and Department of Physics, Chemistry and Biology (IFM), Link\"{o}ping University, 581 83 Link\"{o}ping, Sweden}
\affiliation{Terahertz Materials Analysis Center, THeMAC, Link\"{o}ping University, 581 83 Link\"{o}ping, Sweden}
%\affiliation{Solid State Physics and NanoLund, Lund University, P. O. Box 118, 221 00 Lund, Sweden}

\author{Dat Q. Tran}
\affiliation{Center for III-Nitride Technology, C3NiT-Janz\'en and Department of Physics, Chemistry and Biology (IFM), Link\"{o}ping University, 581 83 Link\"{o}ping, Sweden}

\author{Per O. \AA. Persson}
\affiliation{Thin Film Physics, Department of Physics, Chemistry and Biology (IFM), Link\"{o}ping University, 581 83 Link\"{o}ping, Sweden}

\author{Plamen P. Paskov}
\affiliation{Center for III-Nitride Technology, C3NiT-Janz\'en and Department of Physics, Chemistry and Biology (IFM), Link\"{o}ping University, 581 83 Link\"{o}ping, Sweden}

\author{Vanya Darakchieva}
\affiliation{Center for III-Nitride Technology, C3NiT-Janz\'en and Department of Physics, Chemistry and Biology (IFM), Link\"{o}ping University, 581 83 Link\"{o}ping, Sweden}
\affiliation{Terahertz Materials Analysis Center, THeMAC, Link\"{o}ping University, 581 83 Link\"{o}ping, Sweden}
\affiliation{Solid State Physics and NanoLund, Lund University, P. O. Box 118, 221 00 Lund, Sweden}

\date{\today}% It is always \today, today,

\begin{abstract}
Nitrogen-polar III-nitride heterostructures offer advantages over metal-polar structures in high frequency and high power applications. However, polarity control in III-nitrides is difficult to achieve as a result of unintentional polarity inversion domains (IDs). Herein, we present a comprehensive structural investigation with both atomic detail and thermodynamic analysis of the polarity evolution in low- and high-temperature AlN layers on on-axis and 4$^{\circ}$ off-axis Carbon-face 4H-SiC (000$\bar{1}$) grown by hot-wall metal organic chemical vapor deposition. A polarity control strategy has been developed by variation of thermodynamic Al supersaturation and substrate misorientation angle in order to achieve desired growth mode and polarity. We demonstrate that IDs are totally suppressed for high-temperature AlN nucleation layers when step-flow growth mode is achieved at the off-axis. We employ this approach to demonstrate high quality N-polar epitaxial AlGaN/GaN/AlN heterostructures.

Email: hengfang.zhang@liu.se; vanya.darakchieva@liu.se
%\begin{description}
%\end{description}

\end{abstract}

\maketitle

%\tableofcontents

\section{Introduction}

GaN-based electronic devices are on a rise for high-power and high-frequency device applications needed to enable the transition to green energy and communication systems \cite{Chu_2020GaN, Amano_2018roadmap, Oka_2019recent}. A widely adopted device structure for such applications is the AlGaN/GaN high electron mobility transistors (HEMTs) with a two-dimensional electron gas (2DEG) at the interface \cite{Mishra_2008,Chowdhury_2013,Ted_2018,Ambacher_1999, chang2017strain, baek2021gate}. AlGaN/GaN HEMTs are the most promising devices for millimeter-wave applications such as 5G wireless communications and advanced radars \cite{Mishra_2008, Amano_2018roadmap, zhang2020hetero, zheng2021gallium}. AlGaN/GaN HEMTs provide high breakdown voltage, high carrier mobility and low on-resistance in device structures with а reduced size. Especially, nitrogen-polar (N-polar) HEMTs exhibit unique characteristics compared with metal-polar devices \cite{Koksaldi2018, Romanczyk_2018,Pasayat_2019}, such as the feasibility to fabricate ohmic contact with low resistivity, an enhanced carrier confinement with a natural back barrier, as well as improved device scalability \cite{wong2013n}. However, growth of N-polar group III-nitrides is not trivial because of unintended formation of polarity inversion domains (IDs) \cite{Koukoula_2014self,Hussey_2014,Keller_2006, Fu_2006,Sumiya_2004review}. Notably, the identification of polarity may be challenging by some limitations of commonly used techniques \cite{Zhang_2022}.

Polarity of (Al, Ga)N epilayers and heterostructures is affected by different factors such as V/III ratio, substrate nucleation scheme, impurities etc. For instance, for metal-organic chemical vapor deposition (MOCVD) growth on C-face SiC substrates, where AlN nucleation layers (NLs) were deposited using a low V/III ratio, spontaneous polarity inversion was observed in  N-polar AlN and GaN layers \cite{Keller_2006, Fu_2006}. The pre-flow of trimethyaluminum (TMAl) for surface treatment of C-face SiC substrates was found to result in metal-polar instead of N-polar material \cite{Sumiya_2004review}. Nitrogen-rich growth conditions with a high V/III ratio 
were claimed to be required to realize N-polar III-nitride layers \cite{Keller_2006, Fu_2006}. Furthermore, a distinct unintentional impurity incorporation by oxygen, carbon, and hydrogen in N-polar GaN layers was reported \cite{Fichtenbaum2008}. In our previous study, three-dimensional (3D) growth mode of N-polar AlN NLs on SiC (000$\bar{1}$) was identified as the likely cause for the observed polarity inversion \cite{Zhang_2022}. We suggested that the $\langle$40-41$\rangle$ facets of the 3D islands facilitate enhanced unintentional incorporation of oxygen which leads to the formation of inclined inversion domain boundaries (IDBs) \cite{Zhang_2022}. On the other hand, intentional doping was also found to influence the polarity. For example, pyramidal IDs in Mg-doped GaN epitaxial layers were commonly observed \cite{papamichail_2022Mg, Iwata_2019pyramid, Tetsuo_2018wide}.

Unintentional polarity inversion in N-polar III-nitride heterostructures deteriorates the crystalline quality of the material that further affects negatively the device performance. It is therefore critical to understand polarity evolution and establish strategies to suppress the formation of metal-polar IDs in N-polar layers in order to control the polarity. Specifically, the control of the polarity is essential for N-polar AlGaN/GaN HEMTs \cite{Romanczyk_2018, Pasayat_2019}. Despite the numerous reports on the polarity inversion in III-nitride layers, the underlying physical mechanism of polarity control is still lacking and needs to be addressed. AlN epilayers grown on sapphire substrates are typically N-polar. In order to acquire metal-polar layers, Stolyarchuk $et$ $al.$ \cite{Stolyarchuk_2018} demonstrated an intended polarity inversion from N-polar to Al-polar AlN layers by intentional annealing in oxygen and forming a thin Al$_x$O$_y$N layer at the IDBs. In our previous study, we have reported on growth of N-polar AlN layers on SiC and discussed the effect of substrate orientation on the polarity, surface morphology and crystal quality \cite{Zhang_2020n}. Ponce $et$ $al.$ \cite{Ponce_1996} described atomic arrangements at the interface of AlN on Si-face SiC (0001) from a theoretical perspective and explained how atomic mixing at the interface affects the polarity of AlN. A thermodynamic model was developed to describe the influence of gallium (Ga) supersaturation on the properties of N-polar GaN grown by MOCVD \cite{Mita_2008, Mita_2011}. This model provides a simplified practical approach to analyze the growth of GaN in a non-equilibrium process. The supersaturation can be varied in different way, for example, the choice of diluent gas, V/III ratio, and the growth temperature. Furthermore, Burton, Cabrera and Frank (BCF) model \cite{BCF_1951} can be used to relate (Ga, Al) supersaturation to the surface morphology. Bryan et al. \cite{BRYAN_2016} have extended the BCF model to explain the surface morphology of Al-polar AlN and discussed the dependence of the surface kinetics on the vapor supersaturation and the substrate misorientation angle. These finding indicates that the supersaturation model can be used to predict the surface morphology of III-nitride layers grown by MOCVD.

In this work, we present a comprehensive study on growth of N-polar (Al,Ga)N layers and heterostructures on C-face SiC. First, we examine the growth mode of AlN NLs in relation to polarity evolution by atomically resolved scanning transmission electron microscopy (STEM). A thermodynamic supersaturation model combined with experimental results is utilized to understand the surface morphology of the layers. It is confirmed by the BCF model that the characteristics of the growth surface highly depend on the supersaturation. Furthermore, a polarity control strategy is developed by variation of Ga supersaturation and by tuning the miscut angle of the substrate. Finally, we outline a process that suppresses polarity IDs in N-polar AlGaN/GaN/AlN heterostructures for HEMTs.

\begin{figure*}[t!]
\includegraphics[width=0.6\textwidth]{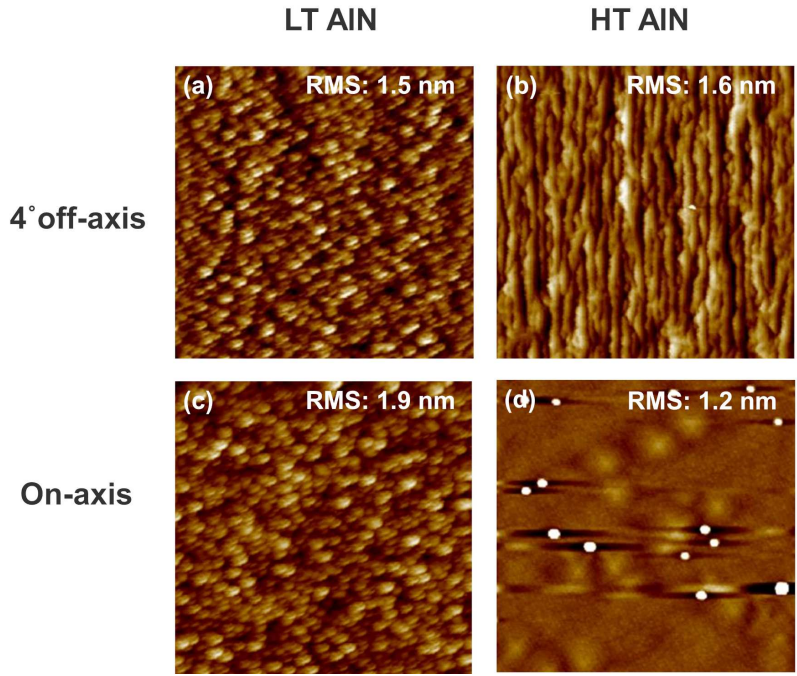}% Here is how to import EPS art
\centering
\caption{3 $\mu$m $\times$ 3 $\mu$m AFM images of LT AlN (a) and HT AlN (b) grown on 4$^{\circ}$ off-axis SiC (000$\bar{1}$), and LT AlN (c) and HT AlN (d) grown on on-axis SiC (000$\bar{1}$).} 
\label{fig:AFM3}
\end{figure*}
 
\begin{table*}
\caption{\label{tab:table1}Summary of samples growth temperature, root-
mean-square (RMS) surface roughness, growth mode and polarity.}
%\resizebox{17cm}{!} {
\begin{ruledtabular}
%\begin{tabular}{p{3cm}p{3cm}p{2cm}p{2cm}p{2cm}}
\begin{tabular}{lcccc}
%\firsthline
%\toprule
%\rowcolor{lightgray}
\textbf{Samples}  & \textbf{Temperature ($^{\circ}$C)}  & \textbf{RMS (nm)}  & \textbf{Growth mode}  & \textbf{Polarity} \\
%\midrule
\hline
LT-AlN on-axis & 1050  & 1.9 & 3D island & Al-polar dominance \\
LT-AlN off-axis & 1050  & 1.5 & 3D island & Al-polar dominance \\
HT-AlN on-axis & 1400  & 1.2 & 2D layer-by-layer & N-polar dominance \\
HT-AlN off-axis & 1300 & 1.6 & Step-flow & Pure N-polar  \\
AlGaN/GaN/AlN off-axis & 1050  & 11.1 & Step-flow & Pure N-polar \\
%\bottomrule
%\lasthline
\end{tabular}
\end{ruledtabular}
%}
\end{table*}
 
\section{Experimental details}

AlN, GaN and AlGaN/GaN/AlN structures were grown in a horizontal hot-wall MOCVD on on-axis semi-insulating (SI) 4H-SiC (000$\bar{1}$) or on \textit{n}-type 4H-SiC (000$\bar{1}$) with a 4$^{\circ}$ misorientation angle towards the [11$\bar{2}$0]. TMAl, trimethylgallium (TMGa) and ammonia (NH$_{3}$) were used as precursors sources for Al, Ga and N, respectively. The pressure in the reactor was kept at 50 mbar. The low-temperature (LT) AlN NLs on on-axis and off-axis C-face SiC substrates were grown simultaneously at 1050 $^{\circ}$C. The high-temperature (HT) AlN NLs was grown at 1400 $^{\circ}$C on on-axis 4H-SiC (000$\bar{1}$) substrate, while for the off-axis 4H-SiC (000$\bar{1}$) substrate the HT AlN NLs was grown at 1300 $^{\circ}$C. The N-polar AlGaN/GaN/AlN HEMT structures was grown by adopting HT AlN NLs on an off-axis 4H-SiC (000$\bar{1}$) substrate. Epi-ready 4H-SiC (000$\bar{1}$) substrates were initially annealed and etched in hydrogen at 1340 $^{\circ}$C. After the pre-flow of NH$_{3}$, LT and HT AlN NLs growth took place with mixed N$_2$ and H$_2$ as the carrier gas for 10 min and a V/III ratio of 800. The growth conditions and the observed growth modes for all samples are summarized in Table~\ref{tab:table1}.

The surface morphology of the layers were investigated by scanning electron microscopy (SEM) and atomic force microscopy (AFM). Zeiss LEO 1550 scanning electron microscope and Bruker AFM Dimension 3100 with tapping mode were used. The polarity of AlN NLs, GaN and AlGaN layers was examined by using KOH with the concentration ~0.18 mol/l at 55  $^{\circ}$C \cite{Guo_2015}. High-resolution X-ray diffraction measurement (HR-XRD) is used in conjunction with KOH etching to characterize and verify the polarity of all layers at macroscopic scale.

STEM was employed to investigate the polarity of LT and HT AlN NLs and the HEMT structure. Characterization was carried out in the image and probe aberration corrected Link\"{o}ping Titan$^3$ 60-300 (S)TEM equipped with a high brightness Schottky field emission gun operated at 300 kV, achieving a resolution of 0.6 \AA. Energy dispersive X-ray (EDX) measurements were performed using a Super-X EDX spectrometer and STEM images were acquired with an annular dark field (ADF) detector. STEM cross-section samples were prepared by standard polishing and argon ion milling procedures. Image processing was carried out using Gatan Microscopy Suite 3.2. Simulations of ADF STEM images were performed using Dr. Probe software \cite{BARTHEL20181} rendering supercells of AlN and adopting standard structure parameters ($a$ = $b$ = 3.11197 \AA, $c$ = 4.98089 \AA, $u_{iso}$ = 0.3869) \cite{Paszkowicz_2004}. Scan images were calculated using a multislice approach for two different polarity IDB structures, namely inclined and horizontal lattice overlap. Microscopy parameters used for the simulations were measured prior to imaging. The parameters include: acceleration 300 kV, temperature 21.5 $^{\circ}$C, convergence semi-angle 21.5 mrad, energy spread 1.3 eV, probe size 0.6 \AA, probe current ~50 pA, 2-fold astigmatism $<$ 2 nm, spherical aberration ~200 nm, chromatic aberration ~1.5 mm, 3-fold astigmatism $<$ 30 nm, axial coma $<$ 30 nm, 4-fold astigmatism ~300 nm, star aberration ~500 nm, ADF angular range ~40-150 mrad. The detector response was measured and included in the simulations. The brightness/contrast levels were set close to nominal values to ensure no saturation and nonlinear intensity transfer occurs. Pixel dwell-time was set to 8 $\mu$s resulting in a 8 s frame time with a $1024\times1024$ detector area. A frozen phonon model was included utilizing 25 modes to account for vibrational effects at room temperature.

\section{Results and discussion}

\begin{figure*}[t!]
\includegraphics[width=\textwidth]{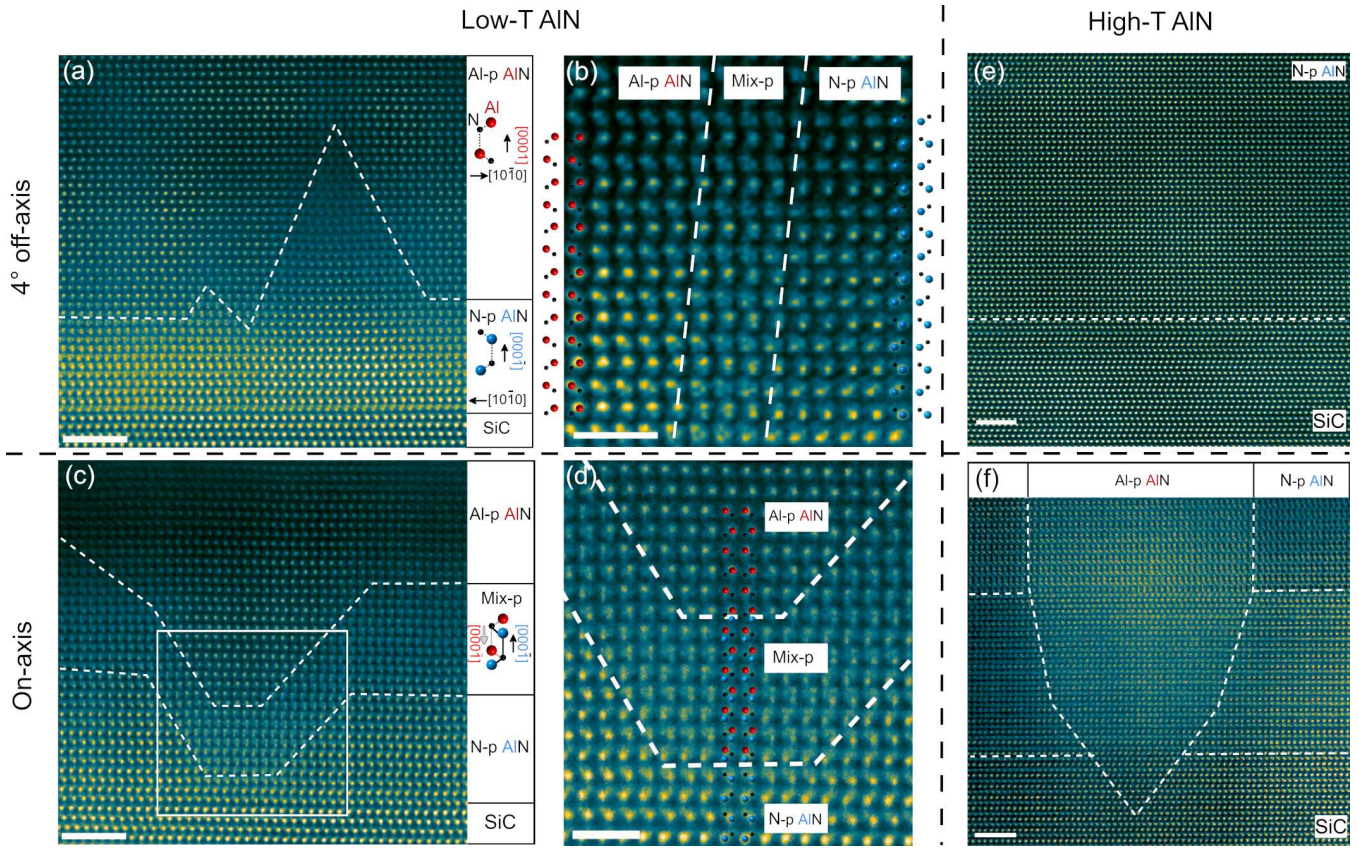}% Here is how to import EPS art
\centering
\caption{LT AlN NLs for 4$^{\circ}$ off-axis (a-b) and on-axis (c-d) growth. (a) Cross-sectional ADF STEM images ([11-20]) showing the off-axis AlN NLs. Regions of different polarities are separated by white dashed lines, the legend illustrates Al-polar (Al-p) and N-polar (N-p) models, in red and blue respectively. (b) Atomically resolved ADF image showing a magnified inclined IDB as observed in (a). (c) Cross-sectional ADF STEM image showing the on-axis AlN NLs. Regions of different polarities are separated by white dashed lines. (d) Atomically resolved ADF image showing inclined and horizontal IDBs as observed in (c). Mix-polar models are shown in overlapping red and blue. Scale bars show 2 nm in (a) and (c), and 1 nm in (b) and (d), respectively. HT AlN NLs for 4$^{\circ}$ off-axis (e) and on-axis (f) growth. Cross-sectional ADF STEM images ([11-20]) showing 4$^{\circ}$ off-axis growth (e) and on-axis growth (f). Atomically resolved ADF STEM image (e) presents a pyramid feature where a mixed-polar lattice is observed. The white dashed line highlights a region with stacking faults. The N-polar growth with an atomically sharp interface to the SiC substrate is also shown. (f) Atomically resolved ADF STEM image verifying an atomically sharp interface between N-polar AlN NLs and SiC showing pyramid-free growth. Scale bars show 2 nm in (e) and (f).}

\label{fig:TEM}
\end{figure*}

\subsection{A. Polarity in LT and HT AlN NLs}

Figure~\ref{fig:AFM3} shows representative AFM images of the surface morphologies of the AlN NLs grown at LT and HT on 4$^{\circ}$ off-axis (a and b) and on-axis (c and d) SiC (000$\bar{1}$) substrates. It is seen that the LT AlN NLs exhibit island growth mode (Figs.~\ref{fig:AFM3} (a) and (c)) regardless of substrate misorientation angle. By using higher growth temperatures of 1300 $^\circ$C for the AlN NLs grown on 4$^{\circ}$ off-axis, step-flow growth mode is achieved (Figs.~\ref{fig:AFM3} (b)). We note that the AlN NLs grown simultaneously at 1300 $^\circ$C on on-axis substrate still exhibits a 3D island growth mode. Therefore, a higher temperature of 1400 $^\circ$C is needed for the AlN NLs on on-axis substrate to transition from 3D growth mode. In this case, 2D layer-by-layer growth mode is observed (Figs.~\ref{fig:AFM3} (d)). The root-mean-square (RMS) surface roughness, or 1.5 nm - 1.6 nm, is similar for AlN NLs on 4$^{\circ}$ off-axis substrates both for island growth and step-flow growth modes. The RMS surface roughness is slightly reduced for the AlN NLs on on-axis substrates when growth mode changes from island growth (1.9 nm, Fig.~\ref{fig:AFM3} (c)) to 2D layer-by-layer growth mode (1.2 nm, Fig.~\ref{fig:AFM3} (d)). The growth temperature of the AlN NLs and their corresponding growth mode, surface roughness, and polarity are summarized in Table~\ref{tab:table1}. 

Notably, the transition from 3D island growth mode for off-axis substrates occurs at lower temperature as compared to on-axis. Typically, on an on-axis substrate, the surface provides large atomically smooth terraces which are much longer than the adatoms diffusion length. 2D nuclei are formed on such a surface, then develop into 3D islands as the epitaxial growth proceeds. On the other hand, for off-axis substrates, the step density increases with a high concentration of kinks on the surface. Therefore, a higher number of favorable nucleation sites are created on the off-axis surface. As a result, the growth mode transition on the on-axis substrate requires a higher temperature (1400 $^\circ$C) to provide sufficient adatom mobility to reach the nucleation sites, while the transition on the off-axis substrate requires a lower temperature (1300 $^\circ$C). When the growth temperature is lower than 1300 $^\circ$C, both on-axis and off-axis exhibit similar 3D island-like growth modes because their surface energy is not sufficient to facilitate long-distance diffusion of adatoms. 

Furthermore, HR-XRD measurements combined with KOH etching indicate that the LT and HT AlN NLs are all N-polar on a macro scale, as evidenced by the disappearance of the AlN 0002 peak upon KOH etching (additional information on polarity determination of samples before and after KOH etching from HR-XRD can be found in the supplementary material \cite{Supplement}). Recently, we have shown that KOH etching may provide inconclusive polarity assignment in case of mix-polarity AlN as Al-polar domains may be under or over-etched through adjacent N-polar pyramid domains, and thus remain undetected \cite{Zhang_2022}. Therefore, cross-sectional atomically resolved ADF STEM is used to determine the polarity by following the stacking order of Al and N atom positions along the growth direction. The LT and HT AlN NLs for on- and off-axis 4H-SiC (000$\bar{1}$) are presented in Fig.~\ref{fig:TEM}. Cross-sectional ADF STEM images on Fig. 2 (a) and (b) show off-axis grown LT AlN viewed along the [11-20] of SiC, while Fig. 2 (c) and (d) present images for the case of on-axis grown LT AlN viewed along the [11-20] as well. White dashed lines on Figure~\ref{fig:TEM} highlight the separation of Al-polar domains close to the surface from mix-polar and N-polar regions at the interface with the SiC substrate. Figure~\ref{fig:TEM} (b) shows a magnified view of the inclined IDB seen in Fig.~\ref{fig:TEM} (a) verifying N-polarity on the left and Al-polarity on the right. In between, a region of overlapping lattices is resolved, which indicates an inclined IDB both laterally in the image but also inclined in the projection over three columns, suggesting a pyramid face at an angle with respect to the electron beam. Figure~\ref{fig:TEM} (c) presents the atomically resolved image of the on-axis grown LT AlN NL. White dashed lines highlights the separation of N-polar and Al-polar domains. The initial AlN layers are N-polar while after a few layers of growth irregular and lateral IDB are observed followed by Al-polar dominating regions. Figure~\ref{fig:TEM} (d) shows a magnified view of the IDB verifying N-polarity in the bottom region and Al-polarity in the top.

For both the on-axis and the off-axis LT AlN NLs, an Al-polar dominated mixed polarity with similar island-like growth mode is observed. The LT AlN exhibits a pyramid structures with an atomic arrangement that can be explained by a lattice overlap of Al- and N-polar domains in projection, i.e. mix-polar regions. Atomic models of N-polar and Al-polar layers are shown in blue and red (Figure~\ref{fig:TEM}), respectively, with the IDB model showing overlapping in the two atomic models. Notably, the LT AlN exhibits steep pyramid shapes for off-axis growth but with a lower density compared to that for on-axis growth (see supplementary material \cite{Supplement}). This observed difference is mainly due to the different substrate misorientation angles as LT AlN NLs are grown simultaneously on on- and off-axis SiC (000$\bar{1}$). Furthermore, it is essential to point out that for both the initial AlN layers are dominantly N-polar. This is the key point why LT and HT AlN NL can be completely etched away at the macro scale using KOH \cite{Zhang_2022}. The observed IDBs are similar to the Al$_{9}$O$_{3}$N$_{7}$ structures reported by Asaka $et$ $ al.$ \cite{ASAKA_2013} and Stolyarchuk $et$ $ al.$ \cite{Stolyarchuk_2018}. We also note that a simple cross-section overlap of adjacent N-polar and Al-polar lattices will result in the same appeared structure (see Fig. S2 in the supplementary material \cite{Supplement} for comparisons of simulated ADF STEM images of different lattice overlap scenarios.) However, such a situation will imply a simultaneous nucleation of Al-polar and N-polar islands on SiC, which are not observed here in any of the TEM images. The IDBs observed in the off-axis AlN NL are of Holt type 2 \cite{HOLT_1969}, and the angle is typical for pyramid shape IDBs previously observed in AlN \cite{Sarigiannidou_2006}. 

In contrast, the growth of HT AlN NLs shows a substantially different situation. Figures~\ref{fig:TEM} (e) and (f) present atomically resolved images of HT AlN NLs grown on off- and on-axis SiC substrates. Figure~\ref{fig:TEM} (f) reveals a large inverted pyramid Al-polar inversion domain surrounded by N-polar AlN regions. However, these IDs have a low density in the film and extend up to the surface, which results in a N-polar dominated mixed polarity (see supplementary material \cite{Supplement}). These inverted pyramidal structures are most likely corresponding to the bright hexagonal structures observed on the surface of the layer (Fig.~\ref{fig:AFM3} (d)). In contrast, the HT AlN NLs on 4$^{\circ}$ off-axis SiC substrate (Fig.~\ref{fig:TEM} (e)) shows a pyramid-free growth with an atomically sharp interface between \textit{pure} N-polar AlN and the SiC. 

These results indicate that there is a connection between growth mode and polarity inversion. The 3D islands growth mode is identified independent of the misorientation angle, and the Al-polar dominated mixed polarity is observed for the LT AlN NLs. On the other hand, 2D layer-by-layer growth mode for the HT AlN leads to N-polar dominated mixed polarity. In contrast, for AlN NLs grown on 4$^{\circ}$ off-axis carbon-faced SiC substrates, the transition of growth modes occurs at a critical temperature of 1300 $^\circ$C, giving a large effect on the correlated polarity. The 3D island growth mode of LT AlN results in N-polar dominated mixed polarity, while the step-flow growth mode of HT AlN results in \textit{pure} N-polarity.

\begin{figure*}[t!]
\includegraphics[width=\textwidth]{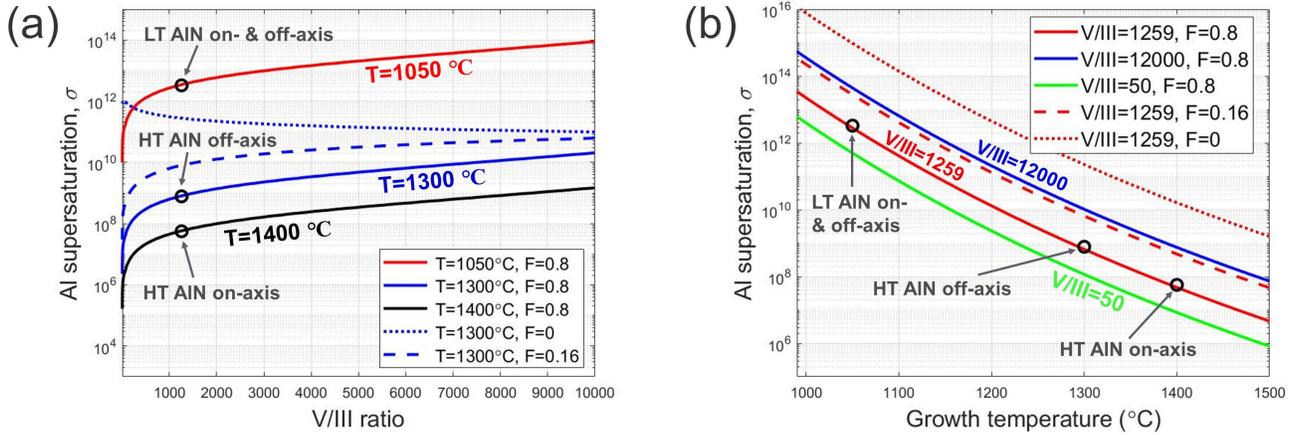}% Here is how to import EPS art
\centering
\caption{Al supersaturation ($\sigma$) for AlN growth versus V/III ratio (a) and growth temperature (b). The calculation was carried out for Al partial pressure of 2.4$\times$10$^{-3}$ mbar. (a) V/III series at different temperatures: 1400 $^\circ$C (black solid line), 1300 $^\circ$C (blue solid line) and 1050 $^\circ$C (red solid line) and different H$_2$ input fractions F: F = 0.8 (solid lines), F = 0.16 (dashed line) and F = 0 (dotted line) a T = 1300 $^\circ$C. (b) Temperature series at different V/III ratios: V/III = 50 (green solid line), V/III = 1259 (red solid line) and V/III = 12000 (blue solid line) and different H$_2$ input fractions F: F = 0.8 (solid lines), F = 0.16 (dashed line) and F = 0 (dotted line) for a V/III = 1259. The black circles indicate the experimental points corresponding to LT and HT AlN on- and off-axis, respectively.}
\label{fig:supersatur1}
\end{figure*}

\begin{figure}[t!]
\includegraphics[width=0.98\columnwidth]{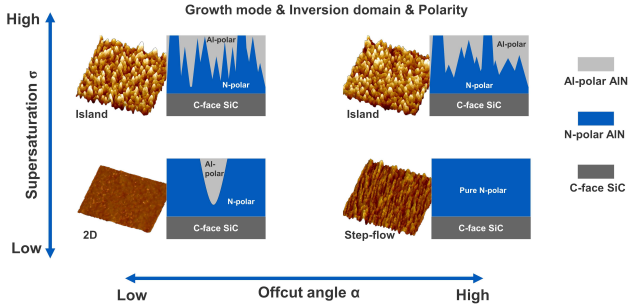}% Here is how to import EPS art
\centering
\caption{A polarity control diagram of AlN NLs via the variation of supersaturation and the misorientation angle of SiC (000$\bar{1}$) substrate. AFM images show different surface morphologies representing different growth modes exhibited in the relevant regions. Corresponding cross-section schematic diagrams are shown with different IDs and polarity. Al-polar AlN is coloured in light grey while N-polar AlN is coloured in dark blue. }
\label{fig:polarity}
\end{figure}

\subsection{B. Growth mode control via Al supersaturation}

In order to understand the surface morphology as a function of growth mode, a thermodynamic model is used to describe the AlN NLs growth. The epitaxial growth is carried out under the mass transport limited condition. In another words, the growth rate is limited by the arriving rate of incoming metal precursor source to the boundary layer between vapour and solid. Growth of epitaxial layers represents a phase transition, where supersaturation plays an important role. The growth of AlN is driven by the supersaturation of Al species in the gas phase transfer to solid crystalline phase under MOCVD environment in a non-equilibrium process. The Al supersaturation is expressed by the relative difference between the input partial pressure of Al and the equilibrium vapor pressure of Al. It is also related to the effect of growth temperature and V/III ratios. Therefore, it could provide important guidance for epitaxial growth from a thermodynamic point of view.

Figure~\ref{fig:supersatur1} (a) shows the calculated Al supersaturation ($\sigma $) as a function of the V/III ratio for different growth temperatures and different H$_2$ mole fractions relative to the total amount of input gases F (F = H$_2$/(H$_2$ + N$_2$)). Similarly, Fig.~\ref{fig:supersatur1} (b) shows the calculated $\sigma $ as a function of growth temperature at different V/III ratios (solid lines: green - V/III = 50, red - V/III = 1259, and blue - V/III = 12000) and different F ratios. The supersaturation values for the four AlN NLs discussed above are shown by black circles in Figs.~\ref{fig:supersatur1} (a) and (b). The calculations are based on the the thermodynamic model proposed by Mita $et$ $ al.$ \cite{Mita_2008} using our standard growth conditions for N-polar growth, Al partial pressure of 2.4$\times$10$^{-3}$ mbar and a total pressure of 50 mbar. N-rich conditions are used and considered in the model. From Fig.~\ref{fig:supersatur1} (a) it is seen that when the V/III ratio is increased up to 10 000, the Al supersaturation increases by about an order of magnitude. In contrast, at a fixed V/III ratio, the increase of growth temperature from 1050 $^\circ$C to 1400 $^\circ$C leads to a significant decrease of Al supersaturation $\sigma$ by about five orders of magnitude. 

The influence of the F ratio on the supersaturation is also shown in Fig.~\ref{fig:supersatur1}, for fixed temperature at varying V/III ratios (Fig.~\ref{fig:supersatur1} (a), blue curves) and for fixed V/III ratio at varying temperatures (Fig.~\ref{fig:supersatur1} (b), red curves). The case of F = 0 represents growth in pure N$_2$ carrier gas. However, this is impossible to achieve in reality since H$_2$ always exists as a byproduct of the growth process from the decomposition of NH$_3$ and the formation of AlN. Therefore, we calculated a small input partial pressure of H$_2$ (F = 0.16, blue dashed curve), e. g. in a N$_2$ rich diluent gas. The Al supersaturation is reduced for lower V/III ratios range. The effect of H$_2$ input fraction on Al supersaturation is less pronounced for the higher V/III ratios (V/III $>$ 5000, blue curves in Fig.~\ref{fig:supersatur1} (a)) as it shows a narrow range of variation. As shown in Fig. 3 (b), from F = 0 (red dotted line) to F = 0.16 (red dashed line) at a fixed V/III ratio, a small F input fraction reduces significantly the Al supersaturation as a function of growth temperature. The Al supersaturation is further reduced as the F ratio is increased, but  not as effectively as for small F values. A similar trend of the Al supersaturation is observed with increasing V/III ratio. Low V/III ratios range shows a more effective influence than the high V/III ratio. It should be noted that the change in F ratios only shifts up and down the curves describing the supersaturation as a function of temperature (red curves in Fig.~\ref{fig:supersatur1} (b)). In contrast, the dependence of supersaturation $\sigma $ on V/III ratios has different trends for different F ratios (blue curves in Fig.~\ref{fig:supersatur1} (a)). 

It is seen from Fig.~\ref{fig:supersatur1} that the LT AlN NLs with 3D growth mode correspond to the growth at high supersaturation. Transition to step-flow mode occurs for off-axis substrate once a supersaturation below 7.7$\times$10$^8$ is reached. A further decrease in supersaturation below 10$^8$ is needed to enable the transition to 2D growth for the AlN NLs grown on on-axis substrate, i.e, 5.6 $\times$ 10$^7$. In our experiments, we have kept the V/III ratio constant and changed the temperature. Arguably, the transitions in growth modes can be achieved by reaching the same critical supersaturation values by changing either V/III ratio, growth temperature or the H$_2$ input partial pressure. For instance, keeping the same F ratio of 0.8, the growth mode transitions may be shifted to lower temperature by reducing the V/III ratio. Alternatively, a further increase of H$_2$ input partial pressure can also be utilized.

\subsection{C. Polarity control strategy}

By the variation of supersaturation and the effect of substrate orientations, we propose a polarity control diagram of AlN NLs grown on SiC (000$\bar{1}$) substrates as illustrated in Fig.~\ref{fig:polarity}. The relations among growth modes, inversion domain types and relevant polarities are presented in this diagram. This diagram also provides a summary of the tendency observed by experiments and explained by the thermodynamic model. This is an important guideline on how to achieve the desired growth mode. The supersaturation value can be adjusted by various growth parameters, such as V/III ratio, growth process pressure, temperature, diluent gas (Fig.~\ref{fig:supersatur1}). In order to obtain pure N-polar AlN, the 3D island growth mode should be avoided which means that a lower supersaturation value should be maintained. We believe that 3D island formation provides multi-planar orientations exposed during growth which further enhanced the probability of oxygen incorporation leading to a polarity inversion \cite{Zhang_2022}. Substrate orientation angle is another key factor to obtain pure N-polar AlN. We found that step-flow growth mode is the crucial point to acquire inversion domain-free layers. We suggest that the step-flow growth mode with pure N-polar could be obtained with the variation of different misorientation angles between 0$^{\circ}$ and 4$^{\circ}$ and the reduction of the supersaturation. 

\begin{figure}[t!]
\includegraphics[width=0.9\columnwidth]{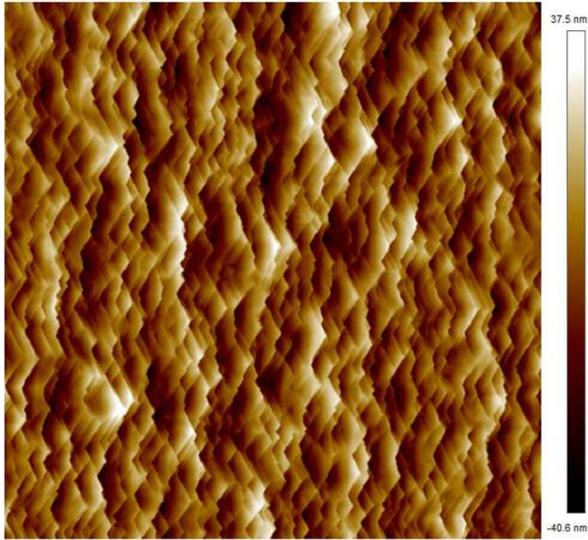}% Here is how to import EPS art
\centering
\caption{AFM image of the top AlGaN surface morphology in the AlGaN/GaN/AlN/SiC HEMT structure. The scanning area is 20 $\mu$m $\times$ 20 $\mu$m with a RMS value of 11.1 nm.}
\label{fig:AlGaN_AFM}
\end{figure}

\begin{figure*}[t!]
\includegraphics[width=0.6\textwidth]{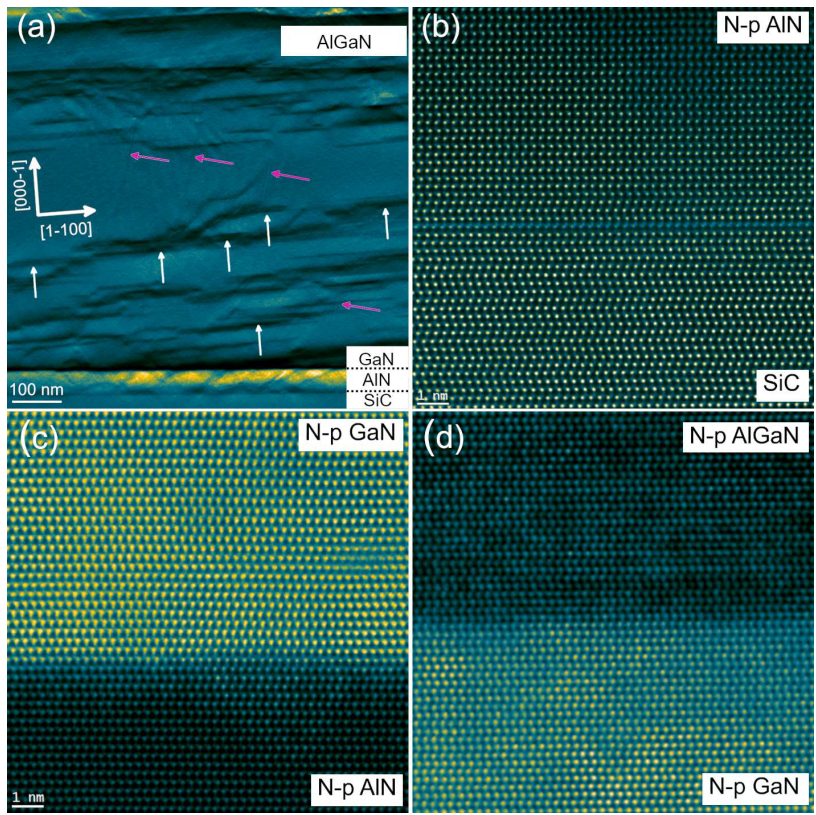}% Here is how to import EPS art
\centering
\caption{(a) Cross-sectional overview ADF STEM image of an AlGaN/GaN/AlN/SiC heterostructure. Atomically resolved ADF STEM images demonstrating the N-polar AlN/SiC interface (b), the GaN/AlN interface (c), and the AlGaN/GaN interface (d).}
\label{fig:AlGaN}
\end{figure*}

\subsection{D. N-polar AlGaN/GaN HEMT structures}

With the polarity control strategy established, we are able to obtain N-polar step-flow growth of AlGaN/GaN/SiC HEMT heterostructures on 4$^{\circ}$ off-axis SiC with a HT AlN NL. Surface morphology of the top AlGaN layer in the HEMT structure is shown in Fig.~\ref{fig:AlGaN_AFM}. The AlGaN/GaN/SiC heterostructures was further investigated by cross-sectional STEM (Fig.~\ref{fig:AlGaN}). The white arrows on Fig.~\ref{fig:AlGaN} (a) indicate stacking faults while the magenta arrows show inclined defects with limited extension. Figures~\ref{fig:AlGaN} (b)-(d) present cross-sectional ADF STEM images identifying the atomic structure and polarity at each interface in the heterostructure -  N-polar AlN and SiC interface (Fig.~\ref{fig:AlGaN} (b)), N-polar GaN and AlN interface (Fig.~\ref{fig:AlGaN} (c)), and N-polar GaN and AlN interface (Fig.~\ref{fig:AlGaN} (d)). Atomically sharp interfaces are obtained which are essential for HEMT device applications. By a further growth optimization, an AlGaN surface roughness of 3.2 nm (on 20 $\mu$m $\times$ 20 $\mu$m scanning area) is achieved for a N-polar HEMT structure. A 2DEG carrier density up to 10$^{13}$ cm$^{-2}$ and an anisotropic electron mobility $\mu_{ex}$/$\mu_{ey}$ = 550/1400 cm$^{-2}$/v$\cdot$s at room temperature has been demonstrated. This relevant detailed study will be reported elsewhere. 

\section{Conclusion}

A comprehensive study in the growth mode and polarity of AlN NLs grown on the C-face SiC substrates is presented. A polarity control strategy for III-nitride layers growth has been developed by variation of (Al, Ga) supersaturation and substrate misorientation angle. It is shown that island growth mode leads to an Al-polar dominated layer for LT AlN NLs regardless of substrate orientation angles. Most importantly, we demonstrate that polarity IDs are completely suppressed for HT AlN NLs when step-flow growth mode is achieved on an off-axis SiC (000$\bar{1}$) substrate. Growth mode control and its relation to the polarity has been explained as a function of supersaturation and substrate orientation angles. We further present a high quality N-polar AlGaN/GaN/SiC heterostructure by effectively suppressing local polarity IDs from forming in the AlN NLs, that ultimately would extend via epitaxy. Our study provides an important step for better understanding of epitaxial N-polar III-nitride layers regarding the surface morphology, growth mode, misorientation angle, polarity inversion, polarity control and their relevant relations. As a result, key guidance for acquiring high-quality device applicable N-polar III-nitride layers are proposed.

\section{Acknowledgement}
This work is performed within the framework of the competence center for III-Nitride technology, C3Nit - Janz\'en supported by the Swedish Governmental Agency for Innovation Systems (VINNOVA) under the Competence Center Program Grant No. 2016-05190, Link\"oping University, Chalmers University of technology, Ericsson, Epiluvac, FMV, Gotmic, Hexagem, Hitachi Energy, On Semiconductor, Saab, SweGaN, UMS and Volvo Cars. We further acknowledge support from the the Swedish Research Council VR under Award No. 2016-00889, Swedish Foundation for Strategic Research under Grants No. RIF14-055 and No. EM16-0024, and the Swedish Government Strategic Research Area in Materials Science on Functional Materials at Link\"oping University, Faculty Grant SFO Mat LiU No. 2009-00971.

\bibliography{apssamp}% Produces the bibliography via BibTeX.

\end{document}